\documentclass{article}

\usepackage[utf8]{inputenc}
\DeclareUnicodeCharacter{202F}{}
\usepackage{textgreek}
\usepackage{amsmath,amssymb}
\usepackage{graphicx}
\graphicspath{ {./images/} }
\usepackage{caption}
\usepackage[superscript]{cite}
\usepackage{comment}
\usepackage{booktabs}
\usepackage{etoolbox}
\usepackage{mathtools, cuted}
\usepackage{lipsum}
\usepackage{IEEEtrantools}
\usepackage{multirow}
\usepackage{xcolor}

\usepackage{parskip}
\usepackage{ragged2e}
\usepackage{hyperref}
\usepackage{textcomp,gensymb}
\usepackage{authblk}

\title{EVIDENCE OF X-RAY EMISSION FROM THE WARM HOT INTERGALACTIC MEDIUM}

\author[1,*]{Giulia Cerini}
\author[2,*]{Nico Cappelluti}
\author[3,*]{Massimiliano Galeazzi}
\author[4]{Eugenio Ursino}
\affil[1,2,3]{Department of Physics, University of Miami, 1320 S Dixie Hway, Coral Gables, FL 33146, USA}
\affil[4]{Department of Physics, Purdue University Fort Wayne, 2101 East Coliseum Boulevard
Fort Wayne, IN 46805, USA}
\affil[*]{Corresponding authors: Giulia Cerini, giulia.cerini@miami.edu; Nico Cappelluti, ncappelluti@miami.edu; Massimiliano Galeazzi, galeazzi@miami.edu.}

\begin{document}

\maketitle
\section{Summary paragraph}
The Universe has evolved from an initial diffuse, uniform gas to a complex structure that includes both voids and high-density galaxy clusters connected by gaseous filaments, known as the Cosmic Web, and traced by 3D surveys of galaxies \cite{Tegmark_2004, Croom_2021}. The filamentary structure contains a significant fraction of the baryonic matter and is predicted to be mostly in the form of a moderately high temperature plasma, the Warm Hot Intergalactic Medium \cite{Cen_1999,Borgani_2004,https://doi.org/10.48550/arxiv.1611.03722}. Plasma at this temperature and ionization level emits mostly in soft X-rays. The filamentary structure, however, is hard to detect because the other sources contributing to the Diffuse X-ray Background are much brighter and, currently, there are very few reported detections of emission from the filaments \cite{Ursino_2006,Werner_2008}. We report the first high-confidence level indirect detection of X-ray emission from the Warm Hot Intergalactic Medium.
Applying the Power Spectrum Analysis to XMM-\textit{Newton} and eROSITA data, we separated its contribution from other sources modeled in previous studies \cite{Helgason_2014,Cappelluti_2012}.
Our result is in good agreement with numerical simulations \cite{2014ApJ...789...55U} and fills a critical gap in the picture of the large-scale structure of the Universe, in which filamentary gas, galaxies and dark matter interact and co-evolve.

\section{Main} \label{sec:main} 
\begin{figure*}[ht]
    \begin{center}
        \textbf{Simulated WHIM}\par\medskip
        \includegraphics[width=1.0\textwidth]{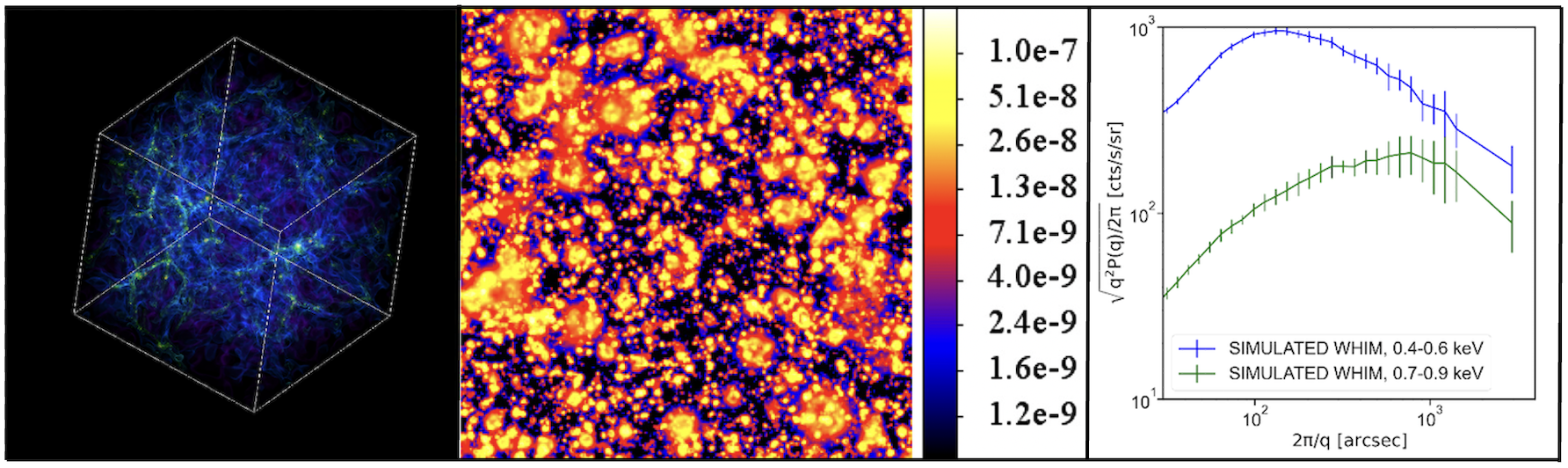}
    \end{center}
    \caption{Detecting the signature of the Cosmic Web in X-rays. Left: starting at a redshift of $z\sim 2$, diffuse baryons in the IGM condense into a filamentary web \cite{2014ApJS..211...19B}. Center: the filaments are not individually visible in X-ray images due to the overlap of many of them along the line of sight; the image represents the simulated X-ray emission, in the band 0.4-0.6 keV, due to the filaments at different redshifts \cite{2014ApJ...789...55U}(units are in $photons\ cm^{-2}s^{-1}$). Right: WHIM signature obtained from the angular power spectrum analysis of simulated X-ray images \cite{2014ApJ...789...55U} (the one in the band 0.4-0.6 keV is shown in the center). The signal peaks around $1\ arcmin$ in the $0.4-0.6\ keV$ energy band, and around $10\ arcmin$ in the $0.7-0.9\ keV$ energy band.}
    \label{fig1}
\end{figure*}

High redshift measurements point to about $4\%$ of the matter-energy density of the Universe to be in the form of baryons, while the rest consists of dark matter and dark energy \cite{Weinberg_1997,Bennett_2003,2020A&A...641A...6P}. In contrast, the amount of baryons measured in the local Universe's structures is less than 2$\%$ \cite{Fukugita_1998,https://doi.org/10.48550/arxiv.astro-ph/0312517,Bregman_2007}. Hydrodynamic simulations tracking the evolution of the Universe in the framework of the commonly accepted $\Lambda$-Cold Dark Matter ($\Lambda CDM$) cosmological paradigm suggest that, starting at a redshift of $z\sim 2$, following the process of structure formation, diffuse baryons in the Intergalactic Medium (IGM) condensed into a filamentary web, the Warm Hot Intergalactic Medium (WHIM), forming the backbone of the Cosmic Web and containing much of the "missing baryons" \cite{Cen_1999,Borgani_2004,Dave_2007,2014ApJS..211...19B,Martizzi_2019} (Fig.~\ref{fig1}-left). These filaments, with a density about 10 to 1000 times the average baryon density $\bar{\rho_b}\sim 10^{-7} cm^{-3}$ (thus ranging from $10^{-6}cm^{-3}$ to $10^{-4}cm^{-3}$), underwent shocks that heated them up to temperatures of $T\sim 10^5- 10^7\ K$, and were enriched with metals through galactic outflows triggered by Supernova and Active Galactic Nuclei (AGN) feedback. At these temperatures and densities, baryons are in the form of highly ionized plasma, making them visible only in the low energy X-ray and UV bands, mostly through excitation lines of highly ionized heavy elements \cite{Ursino_2006}. In the past years UV observations confirmed the presence of the ''warm''  phase of the the intergalactic medium \cite{Tripp_2008,Danforth_2008,Ahoranta20}. However, the study of the ''hot'' phase in the inner regions of the filaments, where the bulk of the missing baryon mass is expected and predicted to emit soft X-rays, has been more challenging. 

In recent years, significant effort has gone towards detecting and characterizing the WHIM, including several ad-hoc missions designed to achieve this goal \cite{10.1117/12.2231193, Barret_2020, 2019JATIS...5b1001G, 2019BAAS...51g.107M,2022arXiv221109827K}. 
Among the methods to detect the WHIM, we highlight the spectroscopic search of absorption along the line of sight of distant bright sources \cite{Kaastra_2006,Fang_2007,Buote_2009,Zappacosta_2010}. However, given the sensitivity of existing X-ray observatories, less than a handful of sources are bright enough to achieve a statistically significant detection. The clearest WHIM detections in absorption were obtained by observing two absorbers of highly ionized oxygen (O VII) in the  X-ray spectrum of the bright X-ray blazar 1ES 1553+113 \cite{Nicastro_2018} with XMM-\textit{Newton}, even if one of them was thought to likely originate from foreground WHIM filaments (indeed, spectroscopic redshift confirmations showed the presence of a spiral galaxy in the proximity of the system, leading to the suspect that the system was actually generated in the group of that galaxy \cite{Johnson_2019}), and studying the emission of the luminous quasar H1821+643 through the combination of Chandra LETG data and previous UV measurements \cite{Kov_cs_2019}. 

Another effort has been done in searching for individual filaments in emission \cite{Werner_2008,Eckert_2015,Sarkar_2022,Veronica_2022} where, however, the inferred temperature of the filament is too high to be considered a true representation of the WHIM. 

\begin{figure*}[ht]
    \begin{center}
        \textbf{DXB components}\par\medskip
        \includegraphics[width=1.0\textwidth]{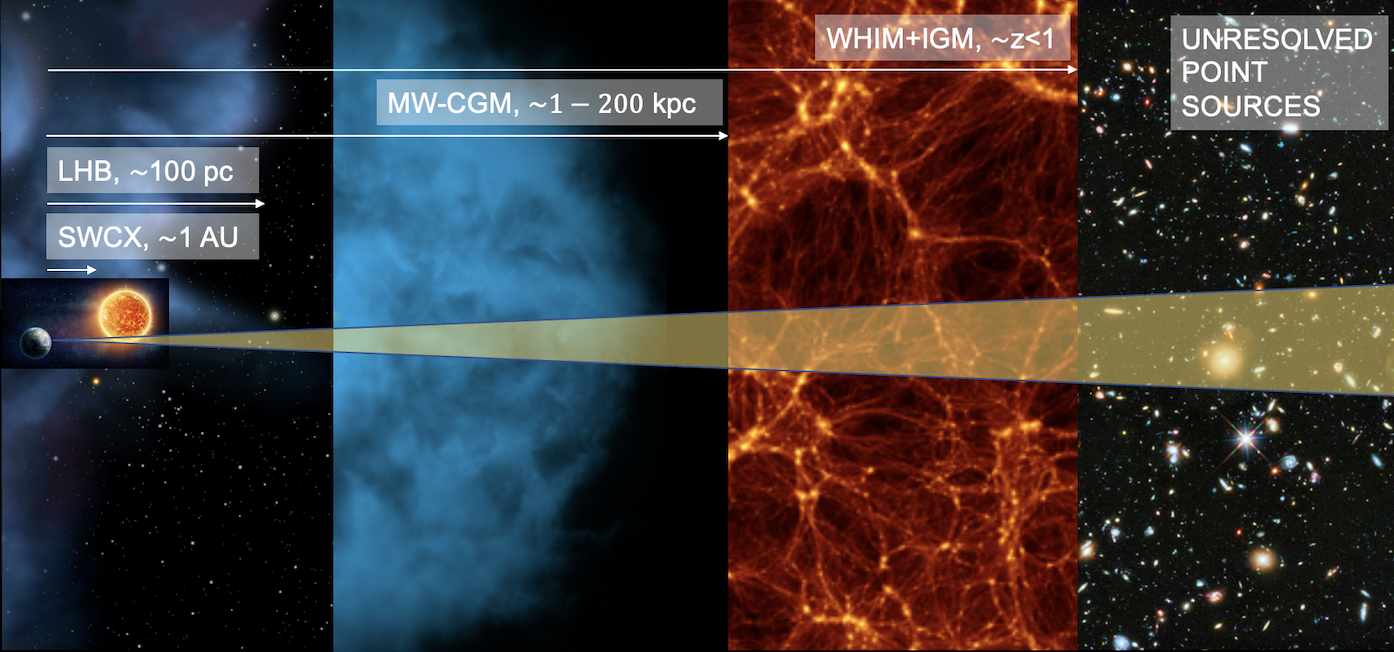}
    \end{center}
    \caption{The X-ray signal from the Cosmic Web filaments is embedded in the foreground and background from other components of the Diffuse X-ray Background. }
    \label{dxb}
\end{figure*}

A third option is to identify redshifted X-ray emission lines emitted by the filaments\cite{Ursino_2006}, but the emission of individual filaments is very weak with respect to other foreground and background X-ray emissions like the Solar Wind Charge eXchange (SWCX), Local Hot Bubble (LHB), Milky Way CGM (MW-CGM), and signal from unresolved point sources, also collectively known as Diffuse X-ray Background (DXB, Fig.~\ref{dxb}). 

In recent years a new method, based on the thermal Sunyaev-Zeldovich effect from the gas filaments, has revealed powerful to detect and investigate the WHIM \cite{de_Graaff_2019,Singari_2020,Hincks_2021}, leading to surprisingly high- confidence level results, even if they do not offer any information about the metallicity of the filaments.


Statistical approaches have also been used to identify and study the filaments. For instance, using XMM-Newton deep fields, the angular auto-correlation function was adopted \cite{Galeazzi_2009} to collect evidence of WHIM emission in the energy band $0.4-0.6\ keV$. More recently,
the use of power spectrum analysis has become the statistical tool of choice for the analysis of diffuse emission at all wavelengths, as it allows to properly disentangle the different components and emitting populations contributing to it \cite{Hickox_2006,Cappelluti_2012,Helgason_2014,Li_2018,Kashlinsky_2018}.
While the current high-resolution X-ray spectrometers are dispersive and have limited effective area, images in narrow energy bands can be generated (Fig.~\ref{fig1}-center). Individual filaments cannot be identified due to the overlap of all the filaments along the line of sight, but the combined signal from all the filaments generates a characteristics angular pattern in the surface brightness fluctuations, whose signature is imprinted in the angular power spectrum. Given the expected temperature and distribution of the WHIM, the signal changes as a function of energy and is significant only below $1\ keV$ \cite{Mushotzky_2000,Giacconi_2001,Ursino_2006}. According to previous studies involving the auto-correlation function in the energy band $0.4-0.6\ keV$ \cite{Galeazzi_2009}, and the power spectrum in the energy band $0.5-2.0\ keV$ \cite{Cappelluti_2012}, the WHIM filaments show a characteristic signal at angles of a few arcminutes. The results from the numerical simulations used in this work \cite{2014ApJ...789...55U} are in agreement with the previous ones, but were obtained in the two narrow energy bands $0.4-0.6\ keV$ and $0.7-0.9\ keV$, and show a peak around $1-10\ arcmin$ that changes with the energy range (Fig.~\ref{fig1}-right). This outcome will be furtherly investigated in follow-up researches, but can be reasonably due to the contribution of filaments at different redshift and projection effects. Indeed, the dominant X-ray flux due to the WHIM is expected to originate from gas at redshift between 0.1 and 0.6 \cite{Ursino_2006}. Photons at higher energies emitted at larger distances are redshifted. Therefore, in the energy band $0.4-0.6\ keV$ we are detecting photons originating at redshift $z\sim0$, as well as photons generated at redshift $z\gtrsim 0.2$, whose energy at the emission was $\gtrsim 0.7\ keV$. On the other hand, the photons in the $0.7-0.9\ keV$ energy range originate from filaments at lower redshifts and, consequently, allow us to see the WHIM in the nearby Universe. The difference in the angular scales involved in statistical approaches like the power spectrum has recently been found also in the redshift evolution of the auto-correlation function obtained from CAMELS hydrodynamic simulations \cite{Parimbelli_2023}. For this particular investigation, the focus was in the two most prominent WHIM emission lines, the OVII triplet and the OVIII singlet, for which line surface brightness maps where built, and 3D and 2D 2-point auto-correlation functions were computed at different redshifts. Their results from the 2D analysis at $z=0.04$ show a plateau at angular scales of size $\lesssim 10\ arcmin$ and fall off in the range $\sim 10-10^2\ arcmin$, while at $z=0.54$ they exhibit a sharper peak at scales of size $\lesssim 1 arcmin$
and a steeper decrease up to $\sim 10\ arcmin$. On the other hand, this difference was not found in the 3D results, that show a pattern more similar to the latter at all redshifts. This was expalained with projections effects that magnify a feature that was difficult to catch in the 3D analysis, namely the core-shaped profile (a similar feature was previously seen also in IllustrisTNG simulations \cite{Nelson_2018}). In addition, the angular scale of the filaments decreases with the increasing redshift, and this contributes to the change of the AcF and, consequently, the power spectrum, as a function of redshift. It is also worth mentioning that the results obtained from the simulations adopted in this work \cite{2014ApJ...789...55U} were demonstrated to be in excellent agreement with the results of other works. Indeed, they showed that, the power spectrum in the energy band $0.4-0.6\ keV$ due to all gas, regerdless of the temperature and density, is in excellent agreement with the values obtained in other works and measured by the Atacama Cosmology Telescope (ACT) \cite{Sievers_2013}, the South Pole Telescope (SPT) \cite{Reichardt2012} and Planck \cite{2014}. The reliability of these simulations was demonstrated in the past also from the analysis of the X-ray properties of galaxy clusters and groups \cite{Borgani_2004}.

To reduce systematics and properly remove the signal from other sources of DXB, 
we used a combination of data from both deep and extended surveys, namely from XMM-Newton (XMM-CDFS, XMM-COSMOS and XMM-Stripe82) and eROSITA (eFEDS - Table \ref{table:1}).  \cite{Ranalli_2013,Comastri_2017,Cappelluti_2009,LaMassa_2013,LaMassa2016,brunner2021erosita}. 
This is important because the DXB source populations show characteristic angular patterns \cite{Cappelluti_2012, Lau22}, with point sources imprinting a clear signal at small scales (of the order of $\sim 10^2\  arcsec$), and diffuse/local sources dominating the largest angular scales.  
With their different Field of View (FoV) and depth (a so-called wedding cake approach) these fields provided the best combination of scales to properly model all the components of the signal and, at the same time, mitigate effects of cosmic variance to achieve high statistical significance.
The study was conducted in 4 energy bands -  $0.4-0.6\ keV$, $0.7-0.9\ keV$, $0.9-1.3\ keV$ and $2.0-5.0\ keV$ (see sec. \ref{sec:methods}) - to look both at the energy ranges of the WHIM signal and in a frequency domain where its contribution is negligible, in order to isolate its signatures. 

In the past decades it has been demonstrated that the signal measured in the power spectrum of the DXB can be analyzed based on population synthesis models and recent observational results, properly combined to predict the amplitude and shape of all its components \cite{Cappelluti_2012,Helgason_2014,Kashlinsky_2018, Lau22}. The amplitude and shape of the power spectrum of the DXB can be interpreted as the superposition of galactic and extragalactic components. Extragalactic components include Shot Noise (SN) from individual undetected AGN and galaxies below the sensitivity of the observation, clustering of galaxies and AGN below the limiting flux \cite{COORAY_2002}, and clustering of undetected  diffuse hot cosmological gas in large scale structures. 
The large scale structure is further divided in the contribution from the WHIM filaments that we are searching for, and hotter/denser structures associated with galaxy clusters and groups (also referred to as IGM). Another component to take into account is the cross-correlation of each individual cosmological population (AGN, galaxies and the IGM) as they share the same environments. 
The galactic contribution comes primarily from the structure of the MW-CGM. Indeed, the emission from the LHB and SWCX can be described as a constant flux on the detector \cite{Cappelluti_2012} and therefore contribute to the Poisson noise only (that has been removed, as explained in Sec. \ref{sec:methods}).

To obtain a ''clean'' signal from the filaments, background flares were removed from XMM-\textit{Newton} and eROSITA observations, and for each energy band, the images were masked to remove resolved point sources and extended X-ray sources, background-subtracted and exposure-corrected (see sec. \ref{sec:methods}). For the same fields and same energy band, all the pointings were co-added in order to produce the final images and calculate the power spectrum for each field, for a total of 16 datasets (4 fields, 4 energy bands). 
For each dataset, after the removal of Poisson noise, point sources, instrumental and cosmic foreground, we checked that no edge effects affected the clean power spectrum \cite{https://doi.org/10.48550/arxiv.2209.06831}, from which we subtracted all the well known components, leaving only the contribution from the MW-CGM and, if present, the WHIM (see Figs. \ref{models46}, \ref{models79}, \ref{models913} and \ref{models2050} in Sec. \ref{sec:extended_data}). While the different fields  have completely different characteristics in terms of solid angle and sensitivity, and despite the data come from two different missions, the resulting power spectra were all consistent with each other, giving us confidence about having removed all the foreground components of the signal in the same way. Fig. \ref{all_sub_models} in Sec. \ref{sec:extended_data} shows the model-subtracted power spectra for all the fields under analysis in all the energy bands under investigation. In order to assess the consistency of the results, for every energy range we performed Two-Sample T-Tests to compare the mean and standard deviation of every pair of data-sets, and check whether or not the average difference between the two measurements is statistically significant from zero. In all cases the T-statistic was lower than the critical value on the test distribution at a confidence level of 99.9$\%$ and, consequently, we failed to reject the Null Hypothesis and established that there is no significant difference between the population means. At this point, for each energy band, the power spectra from all four fields were averaged, using a weighted average with the statistical uncertainty as a weight. 

\begin{figure*}[ht]
    \begin{center}
        \textbf{Mean model-subtracted power spectra}\par\medskip   
        \includegraphics[width=0.99\textwidth]{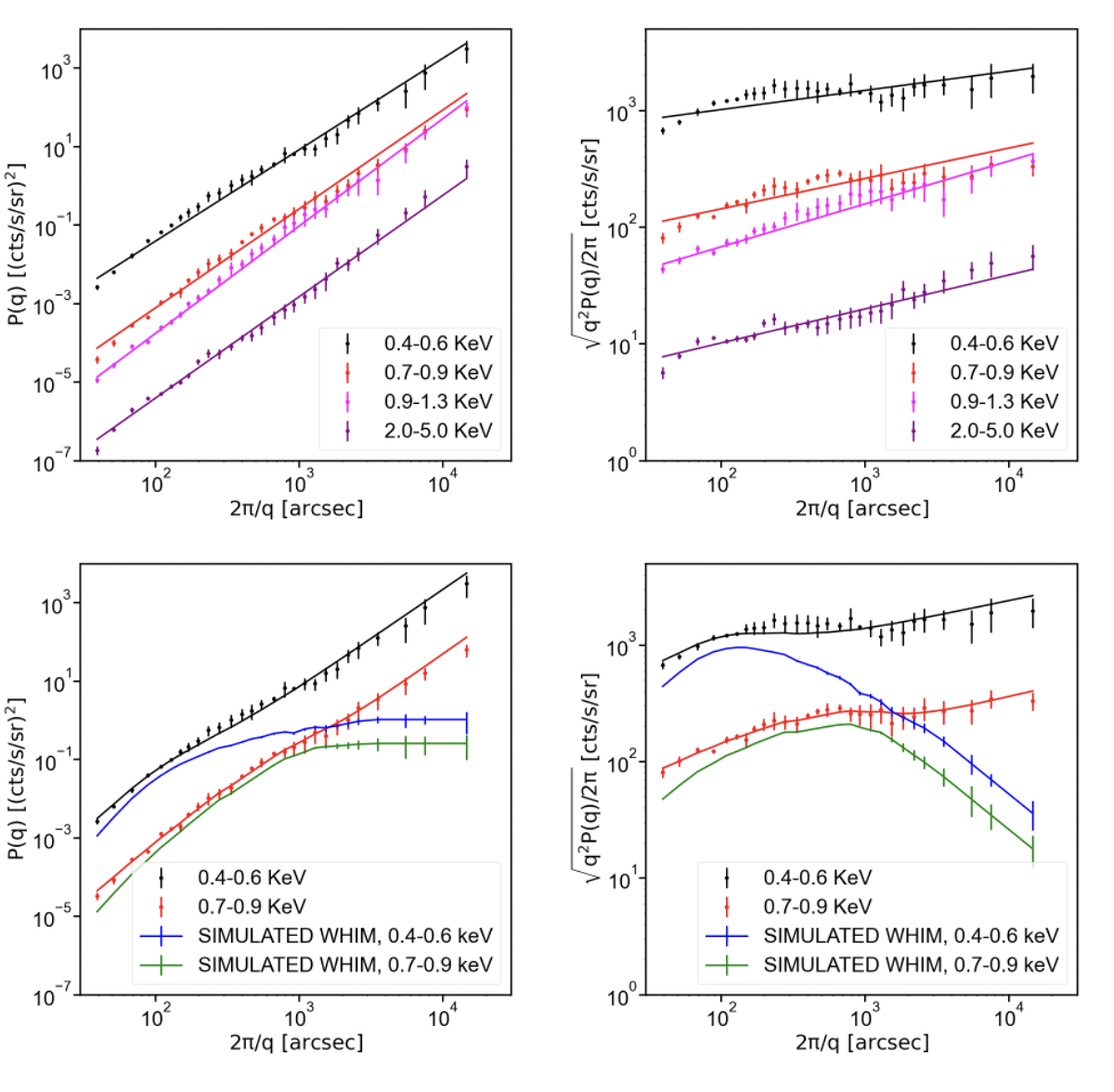}
    \end{center}
    \caption{Mean model-subtracted power spectra obtained averaging, in every energy band, the four model-subtracted power spectra, derived from the analysis conducted in the four fields. The plots in the top row show the best fit lines found adopting a power law $P=a\theta ^k$, while the plots in the bottom row display the best fit lines from the 2-component model $P=a\theta ^k + P_{WHIM}$,  with $P_{WHIM}$ being the blue and green continuous curves with error bars in the plots at the bottom. In both rows, the plots on the left show the power spectra, while the plots on the right contain the RMS fluctuations. The data in the energy bands 0.4-0.6, 0.7-0.9, 0.9-1.3, and 2.0-5.0 keV are shown in black, red, magenta and purple, respectively. The continuous straight lines show the best lines.}
    \label{power_law}
\end{figure*}

Figure \ref{power_law} shows the results for all four energy bands, both in the form of power spectrum (left) and  root mean square RMS fluctuations (right). To analyze the data, we first assume that only one component (the MW-CGM) is contributing. If that is the case, the curve should be fitted with a single power law model $P=a\theta^k$, consistently with what was demonstrated in the past for the galactic cirrus and galactic foregrounds \cite{Bracco,Ingalls_2004,Kashlinsky2005}. As the plots in the top row of Fig. \ref{power_law} show, this is, indeed, true for the energy band $0.9-1.3\ keV$, where the data can be fitted with a simple power law with statistics $\chi^2/d.o.f.=27.57/26$ (see Table \ref{table:2} in sec. \ref{sec:extended_data}). On the other hand, for the energy band $2.0-5.0\ keV$ the statistics $\chi^2/d.o.f.=54.60/26$ suggests that the simple power law fit must be rejected at $3.34\sigma$ confidence level (see Table \ref{table:2} in sec. \ref{sec:extended_data}). This result does not surprise, since the contribution from the MW-CGM to the DXB is still expected at energies $\gtrsim 1\ keV$ but  is negligible at higher energies, and the mean model-subtracted power spectrum is presumably a residual signal for which we cannot find a best fit model. In addition, the value of the exponent of the power law obtained for the $0.9-1.3\ keV$ energy band - $2.68\pm 0.04$ - is in agreement with the typical values in the range $(2,3)$ found for cloud distributions \cite{Bracco,Ingalls_2004,Kashlinsky2005}. It is also worth mentioning that our results are not in conflict with the recent study about the influence of the foreground absorption on the patchiness of the MW-CGM conducted with eFEDS \cite{Ponti2023}. If, indeed, the observed patchiness of the soft X-ray diffuse emission within the eFEDS field is primarily a consequence of absorption, this has important consequences in the spectral analysis, while the angular power spectrum does not change.

The two lower energy bands, $0.4-0.6\ keV$ and $0.7-0.9\ keV$, cannot be similarly fitted with a single power law component (see the plots in the top row of Fig. \ref{power_law}). For these bands, we obtained $\chi^2/d.o.f.=94.64/26$ and $\chi^2/d.o.f.=130.52/26$, respectively. These results suggest that the hypothesis of a single power law model must be rejected at $>6\sigma$ confidence level (see Table \ref{table:2} in sec. \ref{sec:extended_data}).

\begin{figure*}[ht]
    \begin{center}
        \textbf{Whim signal}\par\medskip         \includegraphics[width=0.99\textwidth]{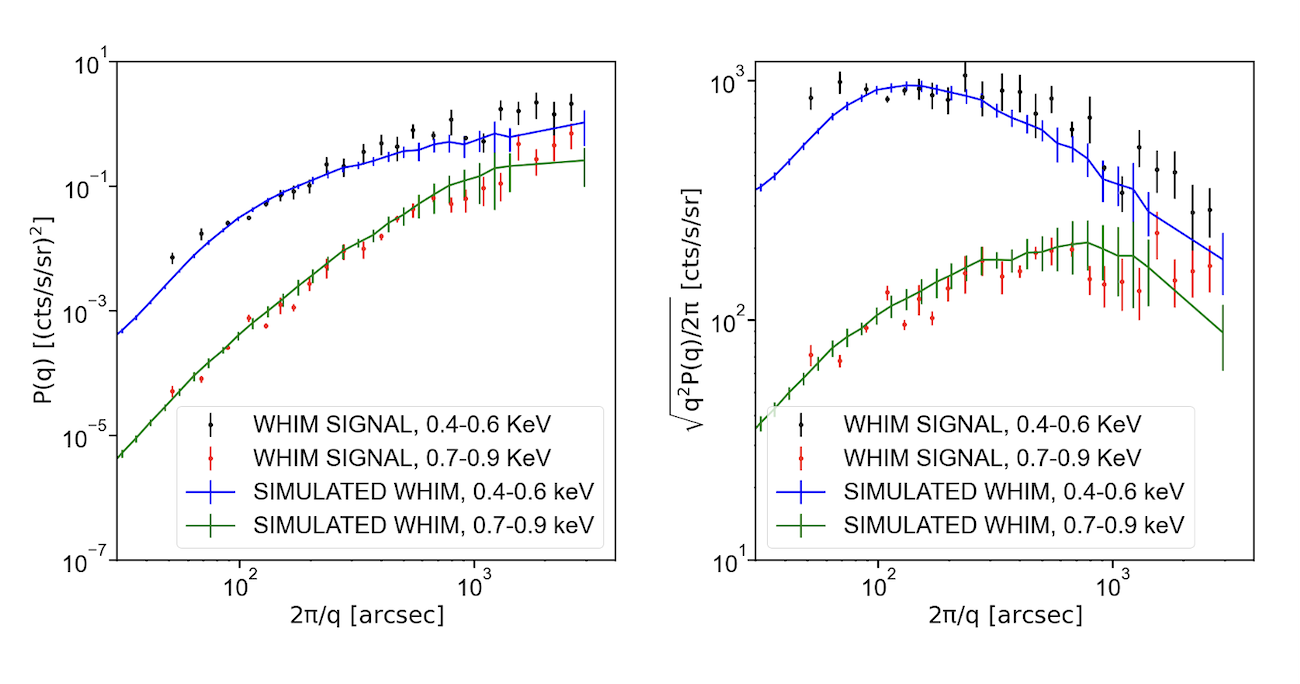}
    \end{center}
    \caption{Both the panels show the observed (black and red data) and simulated WHIM signals (blue and green continuous curves with error bars). The plot on the left contains the power spectrum of the WHIM, while the plot on the right displays the WHIM RMS fluctuations, both obtained after subtracting the best fit power law from the mean model-subtracted power spectra.}
    \label{best_model}
\end{figure*}

\begin{figure*}[h!]
    \begin{center}
        \textbf{Clean power spectra and modeled components in the 0.4-0.6 keV}\par\medskip    
        \includegraphics[width=0.9\textwidth]{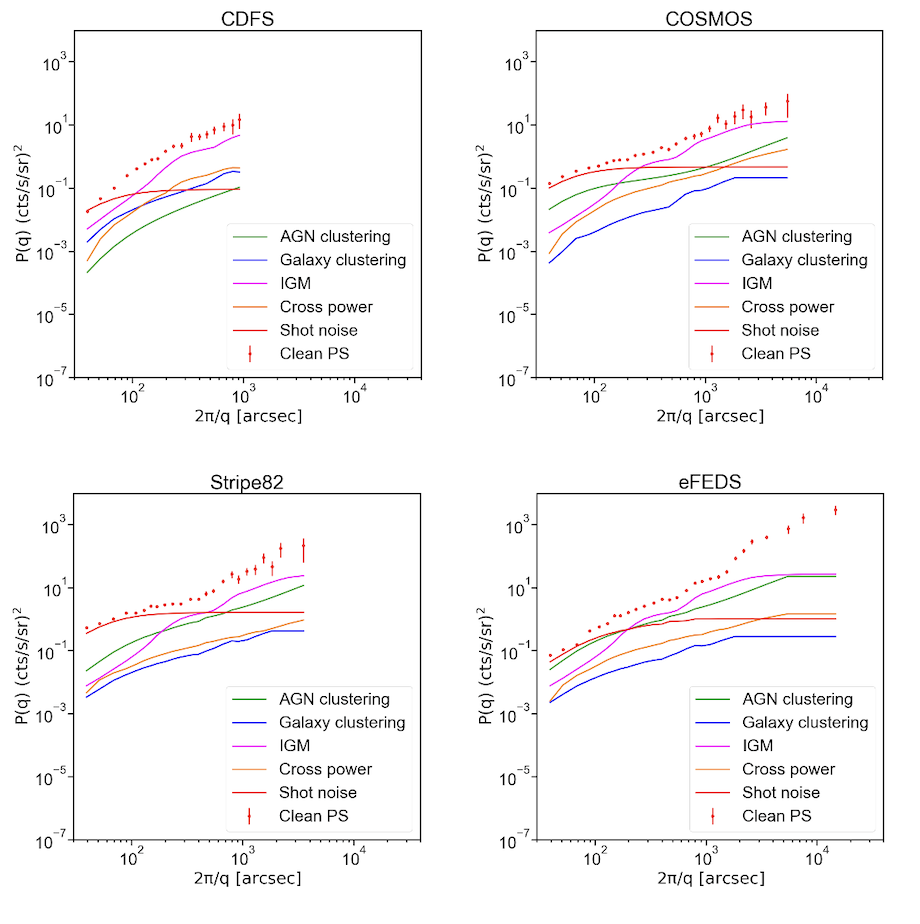}
    \end{center}
    \caption{\textit{Red data}: Clean power spectra for XMM-CDFS \cite{Ranalli_2013,Comastri_2017}, XMM-COSMOS \cite{Cappelluti_2009}, XMM-Stripe82 \cite{LaMassa_2013,LaMassa2016} and eFEDS \cite{brunner2021erosita} in the energy band 0.4-0.6 keV (the data have been obtained after masking the images in all fields, subtracting the background, dividing by the exposure and removing spurious signals). \textit{Red continuous line}: power spectrum of the shot noise. \textit{Magenta continuous line}: power spectrum from unresolved emission from infra-cluster/group gas. \textit{Blue continuous line}: power spectrum due to galaxy clustering. \textit{Green continuous line}: AGN clustering component. \textit{Orange continuous line}: cross-power component.}
    \label{models46}
\end{figure*}

\begin{figure*}[ht]
    \begin{center}
        \textbf{Clean power spectra and modeled components in the 0.7-0.9 keV}\par\medskip     
        \includegraphics[width=0.9\textwidth]{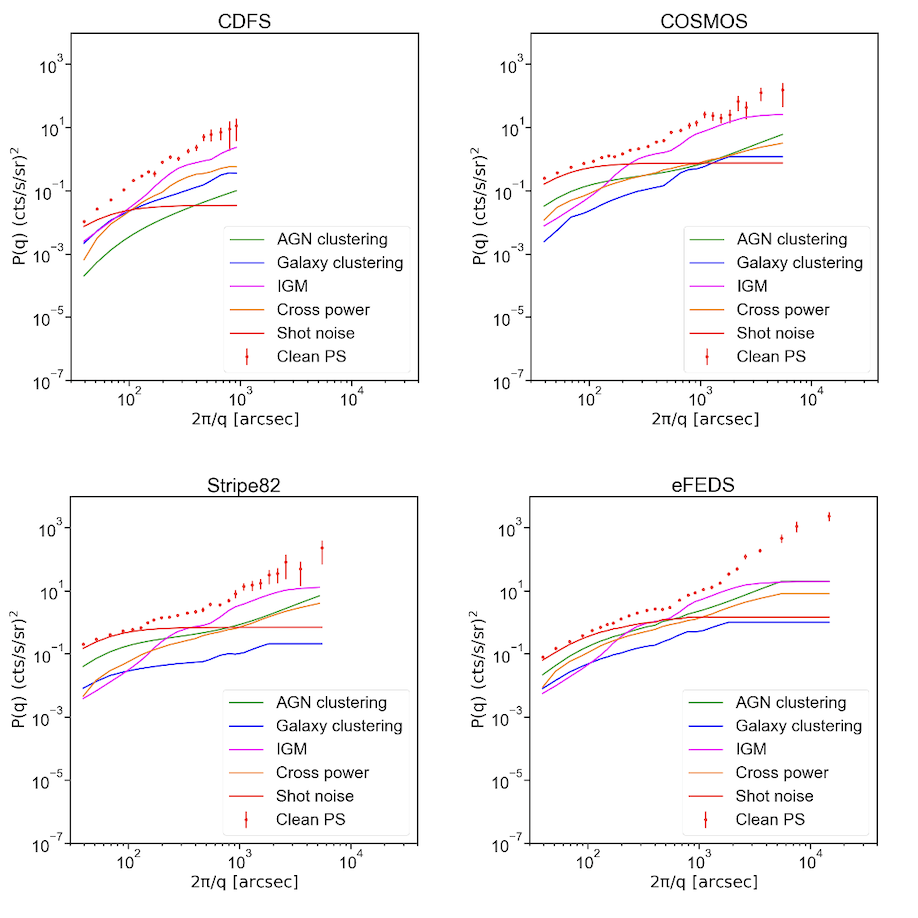}
    \end{center}
    \caption{\textit{Red data}: Clean power spectra for XMM-CDFS \cite{Ranalli_2013,Comastri_2017}, XMM-COSMOS \cite{Cappelluti_2009}, XMM-Stripe82 \cite{LaMassa_2013,LaMassa2016} and eFEDS \cite{brunner2021erosita} in the energy band 0.7-0.9 keV (the data have been obtained after masking the images in all fields, subtracting the background, dividing by the exposure and removing spurious signals). \textit{Red continuous line}: power spectrum of the shot noise. \textit{Magenta continuous line}: power spectrum from unresolved emission from infra-cluster/group gas. \textit{Blue continuous line}: power spectrum due to galaxy clustering. \textit{Green continuous line}: AGN clustering component. \textit{Orange continuous line}: cross-power component.}
    \label{models79}
\end{figure*}

\begin{figure*}[ht]
    \begin{center}
        \textbf{Clean power spectra and modeled components in the 0.9-1.3 keV}\par\medskip     
        \includegraphics[width=0.9\textwidth]{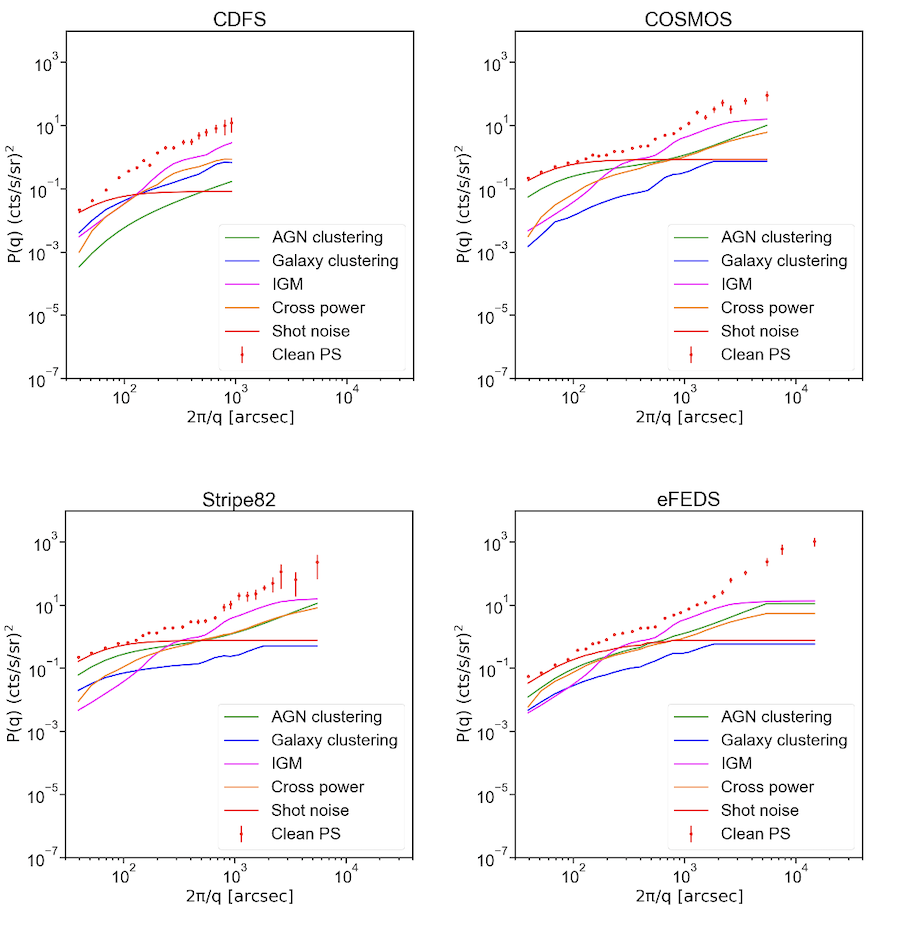}
    \end{center}
    \caption{\textit{Red data}: Clean power spectra for XMM-CDFS \cite{Ranalli_2013,Comastri_2017}, XMM-COSMOS \cite{Cappelluti_2009}, XMM-Stripe82 
   \cite{LaMassa_2013,LaMassa2016} and eFEDS \cite{brunner2021erosita} in the energy band 0.9-1.3 keV (the data have been obtained after masking the images in all fields, subtracting the background, dividing by the exposure and removing spurious signals). \textit{Red continuous line}: power spectrum of the shot noise. \textit{Magenta continuous line}: power spectrum from unresolved emission from infra-cluster/group gas. \textit{Blue continuous line}: power spectrum due to galaxy clustering. \textit{Green continuous line}: AGN clustering component. \textit{Orange continuous line}: cross-power component.}
    \label{models913}
\end{figure*}

\begin{figure*}[ht]
    \begin{center}
        \textbf{Clean power spectra and modeled components in the 2.0-5.0 keV}\par\medskip     
        \includegraphics[width=0.9\textwidth]{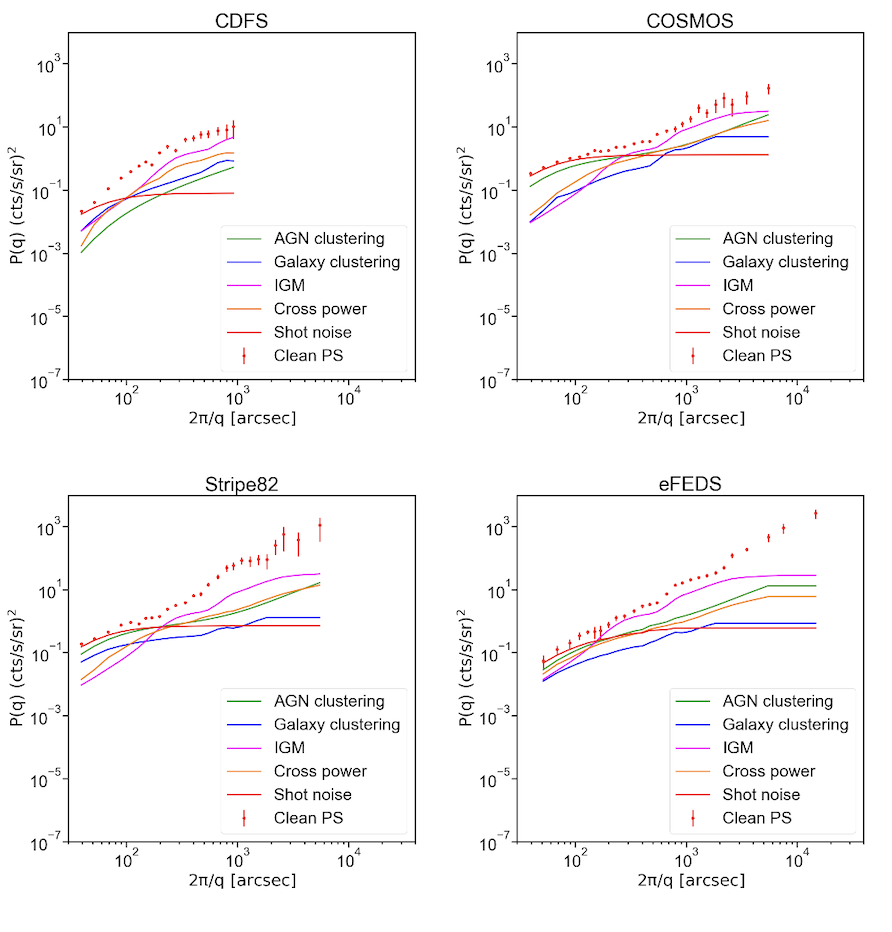}
    \end{center}
    \caption{\textit{Red data}: Clean power spectra for XMM-CDFS \cite{Ranalli_2013,Comastri_2017}, XMM-COSMOS \cite{Cappelluti_2009}, XMM-Stripe82 \cite{LaMassa_2013,LaMassa2016} and eFEDS \cite{brunner2021erosita} in the energy band 2.0-5.0 keV (the data have been obtained after masking the images in all fields, subtracting the background, dividing by the exposure and removing spurious signals). \textit{Red continuous line}: power spectrum of the shot noise. \textit{Magenta continuous line}: power spectrum from unresolved emission from infra-cluster/group gas. \textit{Blue continuous line}: power spectrum due to galaxy clustering. \textit{Green continuous line}: AGN clustering component. \textit{Orange continuous line}: cross-power component.}
    \label{models2050}
\end{figure*}

\begin{figure*}[ht]
    \begin{center}
        \textbf{Model-subtracted power spectra}\par\medskip     
        \includegraphics[width=1.0\textwidth]{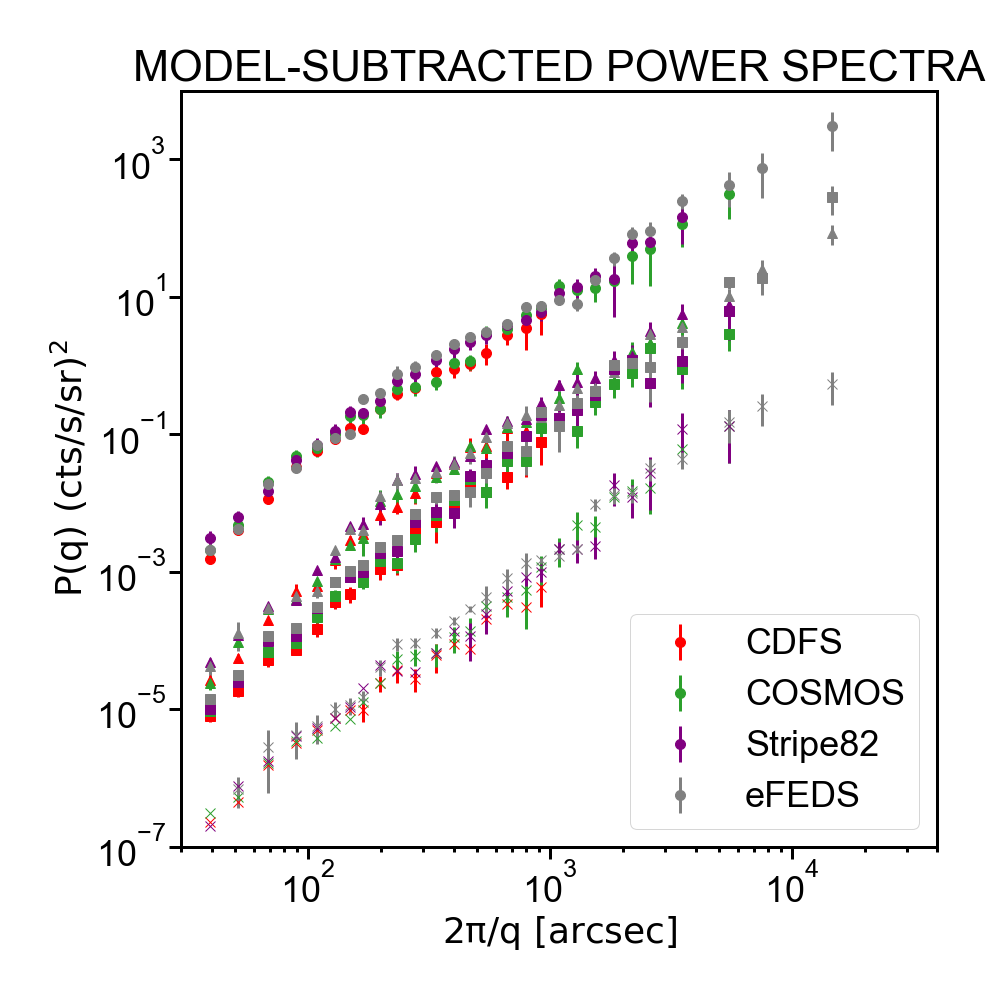}
    \end{center}
    \caption{\textit{Red data}: Model-subtracted power spectra for XMM-CDFS \cite{Ranalli_2013,Comastri_2017}. \textit{Green data}: Model-subtracted power spectra for XMM-COSMOS \cite{Cappelluti_2009}. \textit{Purple data}: Model-subtracted power spectra for XMM-Stripe82 \cite{LaMassa_2013,LaMassa2016}. \textit{Grey data}: Model-subtracted power spectra for eFEDS \cite{brunner2021erosita}. The results in the energy bands 0.4-0.6, 0.7-0.9, 0.9-1.3, 2.0-5.0 keV are shown with circles, triangles, squares and asterisks, respectively.}
    \label{all_sub_models}
\end{figure*}

Up to this point we have shown that, once all other known components have been removed, in addition to the contribution from the MW-CGM, there is a second component that becomes significant below $1\ keV$. To confirm that this component is, indeed, the signature of the WHIM and represents the hot filaments in the Cosmic Web, we combined our data with predictions from cosmological simulations \cite{2014ApJ...789...55U}. The two plots in the bottom row of Fig. \ref{power_law} show the same data we have described before, but this time the model includes two components, a power law to represent the MW-CGM, and the expected WHIM signal derived from cosmological simulations \cite{2014ApJ...789...55U} (Fig.~\ref{fig1}-right). In the fit, while the power law is free to change, the WHIM component has fixed shape and amplitude, and the degrees of freedom have been reduced by 1. 
With the addition of the second cosmological signal, the data in the two lower energy bands are in good agreement with the fit (see Table \ref{table:2} in \ref{sec:extended_data}), with $\chi^2/d.o.f.=30.25/25$ and $\chi^2/d.o.f.=27.00/25$ for the $0.4-0.6\ keV$ and $0.7-0.9\ keV$ bands, respectively. Also for the two lower energy bands we found values of the exponent of the power law due to the MW-CGM consistent with those obtained for the galactic cirrus\cite{Bracco}, namely $k=2.48\pm 0.06$ and $2.54\pm 0.05$, respectively. To compare the two models, in both the lower energy bands we computed $\Delta\chi^2$, $\Delta BIC$ and $\Delta AIC$ (see Table \ref{table:2} in \ref{sec:extended_data}). In all cases the results suggest that
 the model with the WHIM component is favored at $>6\sigma$ confidence level (see Table \ref{table:2} in \ref{sec:extended_data}). So far, even if the spectral resolution of the data does not allow the direct detection of the emission-line signal produced by the WHIM, this result turns to be the highest-confidence level detection of the WHIM in emission. It is also the only high-confidence level signal, so far available, obtained from the combination of data from different and larger regions of the sky, instead of single smaller areas. In Fig. \ref{best_model} we show the WHIM signal, obtained after subtracting the best fit power law from the mean model-subtracted power spectra, compared to the simulated WHIM signal.

\begin{figure*}[ht]
    \begin{center}
        \textbf{Energy spectrum of the WHIM fluctuations}\par\medskip     
        \includegraphics[width=1.0\textwidth]{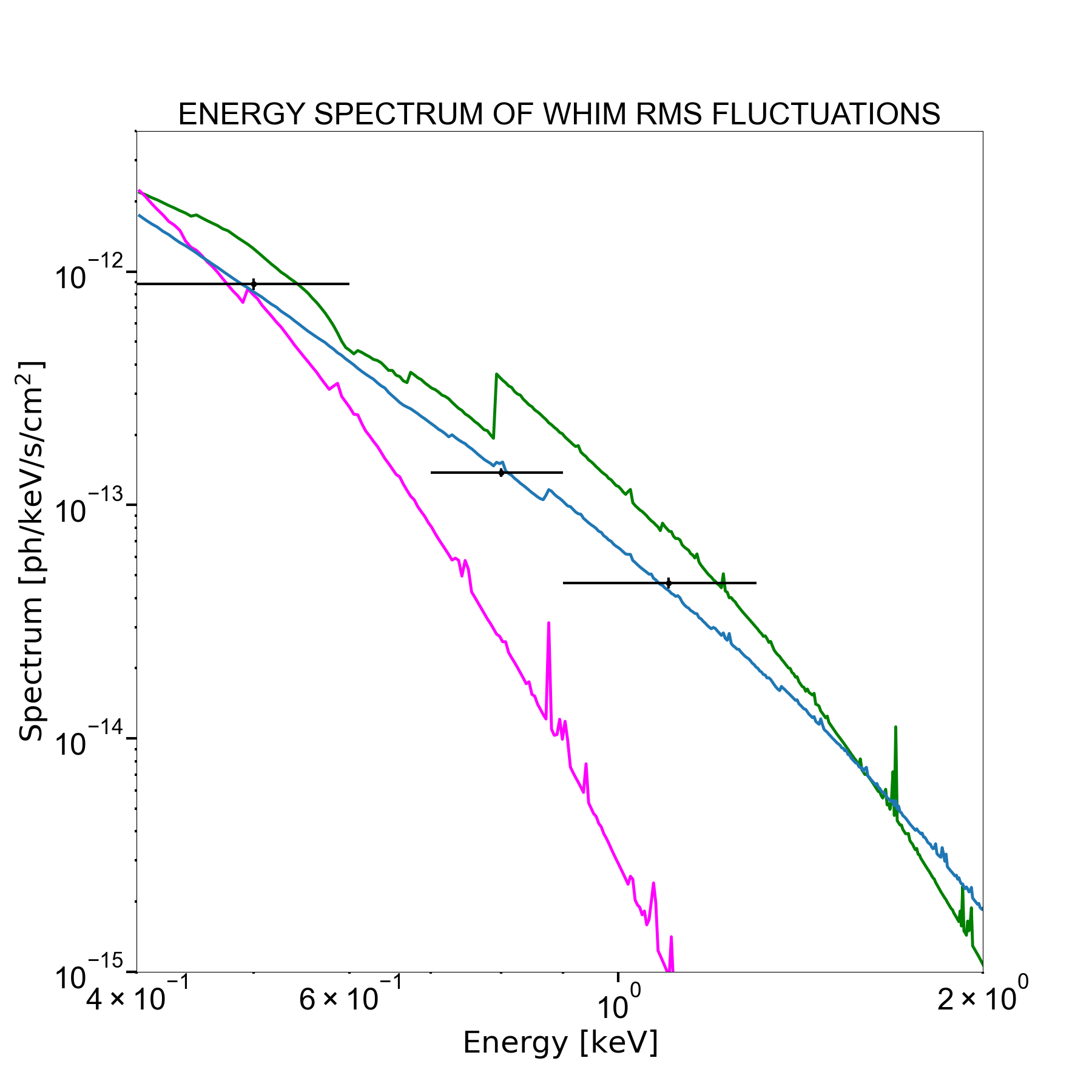}
    \end{center}
    \caption{\textit{Black data}: Energy spectrum of the fluctuations due to the WHIM; \textit{Magenta continuous line}: APEC model with temperature $kT=0.15\ keV$, redshift $z=0.4$ and metallicity $Z=0.01Z_{\odot}$; \textit{Blue continuous line}: APEC model due to the sum of different filaments at redshifts $z=0.0,0.1,0.2,0.3,0.4$, temperatures and metallicities in the ranges $(0.15,0.30)keV$ and $(0.01,0.20)Z_{\odot}$, respectively; \textit{Green continuous line}: APEC model with with $kT=0.25\ keV$, $z=0.0$ and metallicity $Z=0.40Z_{\odot}$. All the normalization values of the APEC models correspond to a gas density $\rho$ in the range $(1,50)\bar{\rho_b}$, through the relation $norm=\frac{10^{-14}}{4\pi[D_A(1+z)]^2}\int{n_en_HdV}$, where $D_A$ is the angular diameter distance to the source, $dV$ is the volume element, $n_e$ and $n_H$ are the electron and H densities, respectively. In particular, for the blue continuous line, we obtained a gas density $\rho\sim10\bar{\rho_b}$. Despite the lack of points to conduct significant statistics, we can notice that the data are included in an area with APEC parameters consistent with the physical properties of the WHIM.}
    \label{e_spectra}
\end{figure*}

Finally, from the RMS fluctuations $\sqrt{q^2P(q)/2\pi}$, shown in all the four energy bands in the right panels of Fig. \ref{power_law}, we reconstructed the energy spectrum of the WHIM fluctuations. To this aim, we picked the value of $\sqrt{q^2P(q)/2\pi}$ at an angular scale of $1.5\ arcmin$ (=$90\ arcsec$) in the three lower energy bands (above $1\ keV$ the contribution from this component is negligible). Plotting $\sqrt{q^2P(q)/2\pi}$ as a function of the energy at the chosen angular scale, (see Fig. \ref{e_spectra} in sec. \ref{sec:extended_data}), we were able to obtain the energy spectrum of the fluctuations due to the WHIM. Using the Simulated Observations of X-ray Sources (SOXS) package, we compared our energy spectrum with different APEC models (Fig. \ref{e_spectra} in sec. \ref{sec:extended_data}), and obtained the following possible ranges for the fitting parameters: $kT\in (0.15,0.30)\ keV$, $\rho \in (1,50) \bar{\rho_b}$, $z \in (0.1,0.4)$, and metallicity $Z\in (0.01,0.40)Z_{\odot}$ (see Fig. \ref{e_spectra} in sec. \ref{sec:extended_data} for more details). These values are all consistent with the physical properties of the WHIM \cite{Borgani_2004, So_tan_2006, Nicastro_2018, Martizzi_2019}.

\newpage
\clearpage

\section{Extended Data} \label{sec:extended_data}

\newpage

\newpage

\newpage

\newpage
\clearpage
\begin{table}
\centering
\textbf{Sample of targets used in this investigation}\par\medskip  
\resizebox{\textwidth}{!}{\begin{tabular}{|l|l|l|l|}
     \hline
\multirow{1}{*}{Target} &
\multirow{1}{*}{Area ($deg^2$)} &
\multirow{1}{*}{Exposure (Ms)} &
\multirow{1}{*}{S$_{lim}(0.5-2)/10^{-15}$  (cgs)}\\
    \hline
 XMM-CDFS & 0.25 & 3.45 & 0.66($2.0-10 keV$)  \\
 \hline
 XMM-COSMOS & 2.13 & 1.5 & 1.7 \\
 \hline
 XMM-STRIPE82 & 20 & 0.8 & 5\\
 \hline
 eFEDS & 140 & 0.36 & 7\\ 
 \hline
 
\end{tabular}}
\caption{Sample of targets used in this investigation. The table entries are: field name (the first column), FoV area (second column), exposure time (third column), sensitivity (fourth column)\cite{Ranalli_2013,Comastri_2017,Cappelluti_2007, Cappelluti_2009, LaMassa_2013, LaMassa2016, brunner2021erosita}.}
\label{table:1}
\end{table}

 \begin{table}
\centering


\textbf{Statistical tests results }\par\medskip
\vspace{1mm}
  \resizebox{\textwidth}{!}{\begin{tabular}{ |l|l|l|l|l|l|l|l|l|l|l|l|l| }
    \hline
    \multirow{2}{*}{Energy band (keV)} &
    \multicolumn{4}{l|}{Model 1, $P=a\theta^k$} &
    \multicolumn{4}{l|}{Model 2, $P=a\theta^k+P_{WHIM}$} & \multirow{2}{*}{$\Delta$BIC} & \multirow{2}{*}{$\Delta$AIC} & \multirow{2}{*}{$\Delta \chi^2$} & \multirow{2}{*}{Confidence} \\ \cline{2-9}
    & $\chi^2$ & d.o.f. & $\chi^2_{red}$ & Confidence & $\chi^2$ & d.o.f. & $\chi^2_{red}$ & Confidence & & & & \\ \cline{2-13}
    \hline
    0.4-0.6 & 94.64 & 26 & 3.64 & $>6\sigma$ & 30.25 & 25 & 1.21 & $1.24\sigma$ & 63.58 & 62.39 & 64.39 & $>6\sigma$ \\
    \hline
    0.7-0.9 & 130.52 & 26 & 5.02 & $>6\sigma$ & 27.00 & 25 & 1.08 & $0.92\sigma$ & 101.61 & 101.52 & 103.52 & $>6\sigma$ \\
    \hline
    0.9-1.3 & 27.57 & 26 & 1.06 & $0.88\sigma$ &  &  &  &  &  &  &  &  \\
    \hline
    2.0-5.0 & 54.60 & 26 & 2.10 & $3.34\sigma$ &  &  &  &  &  &  &  &  \\
    \hline
  \end{tabular}}
\caption{Results of the statistical tests. The table entries are: energy band (first column), results obtained from model 1 (second column), results obtained from models 2 (third column), $\Delta BIC$, $\Delta AIC$, $\Delta \chi ^2$ and $\Delta \chi^2$-obtained confidence level (fourth, fifth, sixth and seventh columns, respectively). For both models, we listed the $\chi ^2$, the d.o.f., the $\chi^2_{red}$ and the confidence level.}  
\label{table:2}
\end{table}

\newpage
\clearpage

\section{Methods} \label{sec:methods}
\subsection{Data assembly} \label{sec:data}
For the purpose of this work, we used the fields listed in Table \ref{table:1} in sec. \ref{sec:extended_data} and numerical simulations \cite{2014ApJ...789...55U}. For the XMM-\textit{Newton} and eROSITA fields, count maps were created in the bands $0.4-0.6\ keV$, $0.7-0.9\ keV$, $0.9-1.3\ keV$ and $2.0-5.0\ keV$, together with background, exposure maps and masks to remove resolved point and extended sources. For every field, the masked, background-subtracted, exposure- corrected images obtained for every pointing were merged and mosaic images were created using the python \textit{reproject.mosaicking} sub-package. The final images have the following FoV: $0.25\ deg^2$ for XMM-CDFS, $2\ deg^2 $ for XMM-COSMOS, $17\ deg^2$ for XMM-Stripe82 and $140\ deg^2$ for eFEDS. The simulated images were obtained in the $0.4-0.6\ keV$ and $0.7-0.9\ keV$ energy band, and have a FoV of $1\ deg^2$.

\subsubsection{XMM-\textit{Newton} data} \label{sec:newton}
For XMM-CDFS, centered at R.A.=$03^h32^m28.2^s$ and dec=$-27^{\circ}48'36.0"$ (J2000),  we used 33 pointings, for a nominal exposure of $3.45\ Ms$ and a total area of $\sim0.25\ deg^2$ \cite{Comastri_2017}. 
The XMM-COSMOS field, centered at R.A.=$10^h00^m28.6^s$ and dec=$+02^{\circ}12'21.0"$ (J2000), covers an area of $\sim2\ deg^2$ and consists of 56 pointings, for a total exposure of $1.5\ Ms$ \cite{Cappelluti_2009}. Finally, for XMM-Stripe82, we combined the 7 observations coming from the $\sim17\ deg^2$ area \cite{LaMassa_2013} of the Announcement Opportunity 13 (AO13), each consisting in 22 pointings, spanning over $14^{\circ}<R.A.<28^{\circ}$, $-0.63^{\circ}<dec<0.63^{\circ}$ (J2000). For all the XMM-\textit{Newton} fields we used data coming from PN and MOS cameras, and performed the reduction of the datasets using the Extended Source Analysis Software (ESAS) package, following the standard data processing procedure for XMM-\textit{Newton} EPIC observations of extended objects and diffuse background \cite{snowden} , and the Science Analysis System (SAS) package.

In order to investigate the DXB, we first removed all resolved sources in the fields. For CDFS we used 59 images (32 PN images and the 27 longest exposure among MOS1 and MOS2) and a combination of 3 catalogs. The first catalog we generated contains 322 XMM-detected point sources using the XMM-SAS tools \textit{emosaic\_prep}, \textit{emosaicproc}, \textit{eboxdetect}, and \textit{region}. Using ds9, we converted the images units to galactic coordinates. For the second catalog, instead, we used the Chandra CDFS-7 Ms source catalog, that reaches a much better sensitivity ($10^{-17}\ erg\ s^{-1}\ cm^{-2}$). It contains the coordinates, redshifts, full-band fluxes, classifications, and many other useful properties for 740 detected sources. Finally, the third catalog  contains 46 extended sources (galaxy clusters and groups \cite{Finoguenov_2015}). The last two catalogs were reformatted to match the XMM-SAS standard and were merged with the
source lists generated by the XMM tools listed above. While the radii of XMM-detected sources are automatically calculated by the XMM tools, to determine the correct cut-out radius of point sources from the Chandra CDFS-7 Ms source and the extended source catalogs , we had to take into account the fact that XMM-\textit{Newton} has a different Point Spread Function (PSF) than Chandra. The appropriate cut-out radius was computed using the linear depeence of the PSF on the off-axis angle \cite{ebrero}. The same approach for the computation of the cut-out radius of resolved point sources was followed for XMM-COSMOS and XMM-Stripe82. For the former, we used the Chandra-COSMOS Legacy Survey Multiwavelength Catalog \cite{Marchesi_2016}, that includes 4016 X-ray sources, 97\% of which have identifications in the optical and Near-Infrared bands. For the latter, the resolved point sources were removed with the Sloan Digital Sky Survey Stripe 82 Chandra Source Match Catalog (SDSS82CXO) and Sloan Digital Sky Survey Stripe 82 XMM-\textit{Newton} Source Match Catalog (SDSS82XMM).

For all the three fields fields, the events registered by the CCDs aboard XMM-\textit{Newton} were cleaned before being processed. One major contamination comes from the Soft Proton (SP) flaring, due to low energy protons entering the telescope from the Van Allen belt. This causes sudden flares in the count rate in the instrument and can lead to a loss of observing time from hundreds of seconds to hours. To remove this contamination we used the XMM-ESAS routines \textit{epchain} and \textit{pn-filter}, that filter the event files from SP flaring and create cleaned events. The cleaned event files were then used by XMM-SAS to create the image count maps with the tool \textit{evselect}. In particular, we sorted events in arrival time to create odd- and even-listed event files, henceforward A and B subsets, from which (A+B)/2 and (A-B)/2 maps were obtained. In particular, the difference between the two subsets does not contain celestial signals or any stable instrumental effects, allowing to evaluate the random noise \cite{Kashlinsky2005,Cappelluti_2013,Cappelluti_2017,Li_2018}. For every field, the A and B substets were created in all the four energy bands. In addition, we generated exposure maps and background maps to subtract other spurious signals, such as cosmic rays and electronic noise \cite{Read_2003}. 

The count maps were then masked, background-subtracted and divided by the exposure to get the count rate maps. The masked, background-subtracted, exposure- corrected images obtained for every pointing were merged and mosaic images were created. In addition, the final count rate maps were divided by the solid angle of the pixels, in order to get the X-ray surface brightness in $counts\ s^{-1}\ sr^{-1}$.

\subsubsection{eROSITA data} \label{sec:erosita}
The eFEDS raw event files were obtained from the eROSITA-DE Early Data Release. These data are made of 4 event files, each corresponding to a 35-square-degree quadralgular patch. The data reduction and analysis was performed using the eROSITA Science Analysis Software System (eSASS) package \cite{2022arXiv220301356C,Klein_2022}. For each event file, the command \textit{evtool} was used to select all photons in the energy bands of interest, converting them into image count maps. To quantify and remove the random noise we used the same A-B subsets approach adopted for the XMM fields. The corresponding exposure maps were created with the \textit{expmap} routine, while the background maps were created combining the tools \textit{erbox} and \textit{erbackmap} \cite{brunner2021erosita}. As for the XMM fields, we subtracted the background maps from the image count maps, that were also divided by the exposure maps and by the solid angle of the pixels, in order to get the same units for the X-ray surface brightness. Resolved point sources and extended X-ray sources were masked out in the X-ray surface brightness maps \cite{2022A&A...661A...3S,Klein_2022}. We used a circular mask to remove every point source at its position. The radius of the circular mask was given by the \textit{ape-radius} parameter in the adopted source catalog \cite{2022A&A...661A...3S}, which is the instrumental PSF with size $3'$. Finally, the four masked, background-subtracted exposure-corrected image count maps were reprojected onto the same World Coordinate System (WCS) and merged together into a single field in all the energy bands under analysis.

\subsubsection{Numerical simulations} \label{sec:simul}
The X-ray images adopted for the evaluation of the WHIM power spectrum come from hydrodynamical simulations \cite{Borgani_2004,2014ApJ...789...55U}, performed with the TREESPH code GADGET-2 \cite{2005MNRAS.364.1105S}, using a $\Lambda CDM$ model with cosmological constant $\Omega_{\Lambda}=0.7$, matter density $\Omega_m=0.3$, baryon density $\Omega_b=0.04$, Hubble constant \(H_0=100\ h\ km\ s^{-1}\ Mpc^{-1}\), $h=0.7$ and $\sigma_8=0.8$. The code follows the evolution of $480^3$ dark matter and baryonic gas particles, from $z=49$ to $z=0$. The physical processes included in the model are gravity, non-radiative hydrodynamics, star formation, feedback from supernovae, taking into account the effects of weak galactic outflows, radiative gas coolong and heating by a uniform, time-dependent, photoionizing ultraviolet background. The output of the simulation consists of 102 $192\ h^{-1}\ Mpc$-side boxes with redshifts between $z=9$ and $z=0$. The metallicity model with the best agreement with constraints from X-ray observations and $Ly\alpha$ observations \cite{Cen_1999} was chosen. The images with the X-ray flux were obtained stacking the simulation boxes in the redshift interval from $z=0$ to $z=3$, then projecting the product of X-ray flux from the single particles to a mesh, using 3-dimensional smoothing kernel as a weight \cite{1985A&A...149..135M}. For every baryonic particle in the simulation, we know density, volume and position, and the metallicity is assigned using a probability distribution function of density and redshift that was previously adopted in other hydrodinamic simulations \cite{Cen_1999b}, from which the X-ray emission can be obtained. The \textit{apec} model for the X-ray Spectral Fitting Package  XSPEC was adopted to generate tables of the energy spectrum as a function of temperature and metallicity. Through an interpolation on temperature and metallicity, and multiplying by $\rho^2$, the X-ray spectrum for every particle can be created and, finally, summing on the redshifted energy bins, the contribution to the bands $0.4-0.6\ keV$ and $0.7-0.9\ keV$ can be computed (the X-ray emission of the single particle was smoothed over the volume elements associated with the particle's position).
The simulation particles were filtered by their density and temperature, in order to obtain images of X-ray emission due to the WHIM, with $10^5<T<10^7\ K$ and $\rho<1000\ \bar{\rho_b}$.

\subsection{Power spectrum formalism} \label{sec:ps}
From the X-ray surface brightness maps, the fluctuation fields \(\delta F_X=F_X(x)-<F_X>\) can be obtained. $F_X(x)$ represents the masked, background-subtracted, exposure-corrected X-ray data, and $<F_X>$ is the average value of the datasets. Through the discrete Fast Fourier Transform (FFT), provided by the python \textit{numpy.fft} subpackage, we computed the Fourier transform 
\(\Delta_X (\textbf{q})\) through the integral \(\int \delta F_X(x)\exp(-i\textbf{x}\cdot\textbf{q})d^2x\), with $\textbf{x}$ coordinate vector in the real space, $\textbf{q}=2\pi \textbf{k}$ wave-vector, $|\textbf{k}|=1/\theta$, and $\theta$ angular scale. The 1D power spectrum $P_X(q)=<|\Delta_X(\textbf{q})|^2>$ can then be computed, where the average is taken over all the independent Fourier elements which lie inside the radial interval $[q,q+dq]$. The error associated to the power spectrum was obtained through the Poissonian estimator $\sigma_{P_X}=P_X(q)/\sqrt{0.5N_q}$, with $N_q/2$ number of independent measurements of $\Delta_i(q)$ out of a ring with $N_q$ data (only one half of the Fourier plane is independent, since the flux is a real quantity). The RMS fluctuations on scales $\theta=2\pi /q$ are usually computed as $\sqrt{q^2P(q)/2\pi}$, where $P(q)$ is the 1D power spectrum of the image under analysis. As previously mentioned, in order to quantify the random noise, we created signal maps $1/2(A+B)$ and noise maps $1/2(A-B)$. The final, clean power spectrum $P_X$ was evaluated as $P_{1/2(A+B)}-P_{1/2(A-B)}$. Since $P_{1/2(A-B)}$ is the power spectrum of fluctuations due to Poisson noise, it is essentially a flat spectrum (white noise). 

\subsection{Modeling the power spectra of the DXB fluctuations} \label{sec:model}

\subsubsection{Shot noise and Point Spread Function} \label{sec:shot}
The shot noise (SN) is produced by discrete sources that occasionally enter in the beam. Indeed, apart from the bright detected sources, that were removed from the images, there do exist faint objects that are not detectable and yet add a further noise component to the power spectrum. Their contribution can be estimated starting from the Log N-Log S and limited fluxes. It has been shown that \cite{Kashlinsky2005}, if the resolved sources are removed down to a flux $S_{lim}$, the shot noise component can be evaluated as $P_{SN}=\int_{0}^{S_{lim}} S^2\frac{dN_X}{dS}dS$, where $dN_X/dS$ is the differential Log N-Log S \cite{Lehmer_2012} of all the X-ray point sources - namely AGN, galaxies and stars. Since the sensitivity of the survey is usually inhomogeneous accross the FoV, $S_{lim}$ is a function of the sky coordinates. To take into account this effect, for every field we scaled the number counts by the corresponding selection functions $\eta(S)$ \cite{Cappelluti_2009,Cappelluti_2012,LaMassa2016,brunner2021erosita}: $   \frac{dN_X}{dS}_{(SN)}=(1-\eta (S)) \frac{dN_X}{dS}$. 

\subsubsection{Point Spread Function} \label{sec:psf}
All the astronomical components of the power spectrum are affected by the telescope (PSF), whose effect consists in a multiplicative factor applied to the effective power spectrum. For the XMM-\textit{Newton} fields we modeled the effect due to the PSF multiplying at every frequency \cite{Churazov_2012} the power spectrum of each modeled component by the factor $P_{PSF}(k)=1/[1+(k/0.02)^2]^{1.6}$, with $k$ angular frequency. For eFEDS, instead, we simulated an image of randomly distributed X-ray point sources with the SImulation of X-ray TElescopes (SIXTE) software package \cite{Dauser_2019} and computed its power spectrum. The power spectrum of randomly distributed point sources is essentially a flat spectrum, while the telescope PSF makes the power spectrum fall off at scales $\theta < \sigma$ (or, equivalently, wave numbers $k>1/\sigma$), where $\sigma$ is the Gaussian width of the telescope PSF. The power spectrum of the PSF can be modeled normilizing to 1 the power spectrum obtained from the simulated image. The SIXTE software simulated the image of point sources generating a number of photons that then are propagated through the representation of the optics of the eROSITA telescope, resulting in a list of impact times, positions and energies. Finally the read-out and the event list are simulated. 

\subsubsection{AGN and galaxy clustering, intergalactic medium and cross correlation terms} \label{sec:agn_gal_cl}
The AGN (and, similarly, the galaxy) clustering is related to the unresolved DXB production rate $dS/dz$, with $S$ the X-ray flux and $z$ the redshift, and the evolving 3D power spectrum of the AGN $P_{3,AGN}(q)$, through the Limber's equation \cite{1980lssu.book.....P}. The power spectrum due to AGN can be related to the matter power spectrum through the redshift dependent linear biasing factor \cite{1984ApJ...284L...9K}: $P_{3,AGN}(k,z)=b(z)^2P_{3,M}(k,z)$, with $P_{3,M}(k,z)$ 3D matter power spectrum. The latter can be estimated using the tool CAMB 
\footnote{Source:\href{http://camb.info/}{http://camb.info/}}
\cite{PhysRevD.66.103511}. The 2D power spectrum can then be obtained with $P_{2,AGN}(k)=\int_{0}^{z} (\frac{dS}{dz})^2\times \dfrac{P_{3,AGN}(k[2\pi d_A\times (1+z)]^{-1}),z}{c\ dt/dz\ [d_A\times (1+z)]^2} \frac{dz}{1+z}$, where $d_A$ is the angular diameter distance. For the Galaxy clustering the procedure is similar. In both cases the flux originating from undetected sources $\frac{dS}{dz}$ can be obtained from DXB synthesis models \cite{Cappelluti_2012} and can be computed as $\frac{dS}{dz}=\int_{0}{\infty} (1-\eta(S))\ \int_{z}{z+dz}\frac{L'}{4\pi d_L^2}\phi(L',z)\frac{dV}{dz}dL'dz$, with $d_L$ luminosity distance, $L'$ luminosity measured in the energy bands $0.4(1+z)-0.6(1+z)$ keV, $0.7(1+z)-0.9(1+z)$ keV, $0.9(1+z)-1.3(1+z)$ keV and $2.0(1+z)-5.0(1+z)$ keV, and $\frac{dV}{dz}$ comoving volume element. \cite{Cappelluti_2012}.

From the analysis of the unresolved DXB from the Chandra observations in the CDFS \cite{Cappelluti_2012}, it was found that another significant contribution to the total DXB signal arises from clustering of unresolved galaxy clusters and groups. In this work we followed the same approach and the same hydrodinamical simulations \cite{Roncarelli2012} to model this component, namely the X-ray surface brightness originating by the intergalactic gas at temperature $T>10^7\ K$, and density $\rho>10^3\rho_c$. We also adopted the same procedure for the computation of the cross-correlation terms. 

\section{Data availability}
The entire XMM-\textit{Newton} and eROSITA data used in this work are available in the public XMM-Newton and eROSITA archives, namely, the XMM-Newton Science Archive (http://nxsa.esac.esa.int/nxsa-web/) and the eROSITA Early Data Release (EDR) archive (https://erosita.mpe.mpg.de/edr/). In particular, we used all the 33 pointings of XMM-CDFS, the 56 pointings of XMM-COSMOS, and the 7 observations of the $\sim\ 17\ deg^2$ AO13 Stripe82 area, each one consisting of 22 pointings \cite{LaMassa_2013}. We also used all the 4 event files corresponding to the 35-square-degree quadralgular patches composing eFEDS. 

\section{Code availability}
The codes for the simulations \cite{Borgani_2004,2010ApJ...721...46U,Roncarelli2012} can be accessed at https://wwwmpa.mpa-garching.mpg.de/gadget/, while the power spectrum analysis and the models of the DXB components used in this work are available upon request from the corresponding authors. The SOXS package for the simulated X-ray observations of astrophysical sources, used for the APEC model, is available at https://hea-www.cfa.harvard.edu/soxs/.

\section*{Acknowledgments}
This research has made use of the SIXTE software package \cite{Dauser_2019} provided by ECAP/Remeis observatory (https://github.com/thdauser/sixte). In particular, we deeply acknowledge Sicong Huang for providing us the XMM-COSMOS and XMM-Stripe82 images, as well as Stefano Marchesi for his support on the SIXTE simulations.

\section*{Author contributions}
Giulia Cerini developed all the codes for the power spectrum analysis, reduced the XMM-COSMOS and XMM-Stripe82 data images, elaborated, together with Nico cappelluti, the pipelines to model all the DXB components, extracted the WHIM signal, and conducted the statistical tests to assess the significance of the results. Nico Cappelluti developed the theoretical framework and techniques for the decomposition of the DXB in its components, and reduced the eFEDS images. Massimiliano Galeazzi designed the project suggesting the analysis in narrow energy bands. Eugenio Ursino provided the numerical simulations, reduced the XMM-CDFS images and, together with Massimiliano Galeazzi, conducted the preliminary test of the method using the auto-correlation function. All authors contributed to the the interpretation of the power spectra, gave theoretical support for the results, and commented on the manuscript.

\subsection*{Corresponding authors}
Giulia cerini, Nico Cappelluti, Massimiliano Galeazzi

giulia.cerini@miami.edu, ncappelluti@miami.edu, galeazzi@miami.edu

\section*{Ethics declaration}
The team has a strong history of supporting diversity and fostering an inclusive workplace to ensure all current and prospective members of the organizations are valued for the unique contributions they bring to support program requirements at their highest levels of performance, and are committed to the vision represented by NASA's core value of inclusion in project activities throughout the full life cycle. At the institutional level, Diversity and Inclusion efforts fall into three main categories: bring practices, community outreach, and training. Both the University of Miami and Purdue University at Fort Wayne advance their goals through regular training, issuance of guidelines to all partner institutions, regular collection and reporting of metrics, issuance of anonymous surveys, and the establishment of well-defined lines of authority to align practices with diversity and inclusion objectives.
\subsection*{Competing interests}
The authors declare no competing interests.

\bibliography{whim_paper}{}
\bibliographystyle{naturemag}
\end{document}